\documentclass{amsart}
\usepackage[english]{babel}
\usepackage[latin1]{inputenc}
\usepackage[dvips,final]{graphics}
\usepackage{amsmath,amsfonts,amssymb,amsthm,amscd,array,stmaryrd,mathrsfs}
\usepackage{pstricks}
 \usepackage[all]{xy}
\usepackage{textcomp}
 \usepackage[final]{epsfig}

\theoremstyle{plain}
\newtheorem{thm}{Theorem}
\newtheorem{lem}{Lemma}[section]
\newtheorem{cor}[lem]{Corollary}
\newtheorem{prop}[thm]{Proposition}

\theoremstyle{definition}
\newtheorem{defi}[lem]{Definition}

\newtheorem{ex}[lem]{Example}

\newcommand{\R}{\mathbb{R}}
\newcommand{\Z}{\mathbb{Z}}
\newcommand{\C}{\mathbb{C}}

\newcommand{\bbH}{\mathbb{H}}
\newcommand{\bbK}{\mathbb{K}}

\newcommand{\K}{\mathbb{K}}

\newcommand{\A}{\mathcal{A}}

\newcommand{\Cm}{\mathcal{C}}

\newcommand{\Cl}{\mathrm{Cl}}
\newcommand{\M}{\mathrm{M}}

\newcommand{\e}{\varepsilon}

\def\a{\alpha}
\def\b{\beta}

\def\e{\varepsilon}

\def\g{\gamma}
\def\G{\Gamma}

\begin{document}

\title{Graded commutative algebras:\\
examples, classification, open problems}

\author{Sophie Morier-Genoud
\hskip 1cm
Valentin Ovsienko}

\address{
Sophie Morier-Genoud,
Universit\'e Paris Diderot Paris 7,
UFR de math\'ematiques case 7012,
75205 Paris Cedex 13, France}

\address{
Valentin Ovsienko,
CNRS,
Institut Camille Jordan,
Universit\'e Claude Bernard Lyon~1,
43 boulevard du 11 novembre 1918,
69622 Villeurbanne cedex,
France}

\email{sophiemg@math.jussieu.fr,
ovsienko@math.univ-lyon1.fr}

\date{}


\begin{abstract}
We consider $\G$-graded commutative algebras,
where $\G$ is an abelian group.
Starting from a remarkable example of the classical
algebra of quaternions and, more generally, an arbitrary Clifford algebra,
we develop a general viewpoint on the subject.
We then give a recent classification result and formulate an
open problem.
\end{abstract}

\maketitle

\thispagestyle{empty}



\section{Introduction}

\subsection*{The algebra of quaternions}

Our first example of graded commutative algebra
is the classical algebra of quaternions, $\bbH$.
This is a 4-dimensional associative algebra with the basis
$\{1,i,j,k\}$ and relations expressed by the
celebrated formula of Hamilton:
$$i^2=j^2=k^2=i\cdot{}j\cdot{}k=-1.$$
It turns out that $\bbH$ is commutative in the following sense.
Associate the ``triple degree'' to the basis elements:
\begin{equation}
\label{DegH}
\begin{array}{rcl}
\bar{\e}&=&(0,0,0),\\[4pt]
\bar{i}&=&(0,1,1),\\[4pt]
\bar{j}&=&(1,0,1),\\[4pt]
\bar{k}&=&(1,1,0),
\end{array}
\end{equation}
viewed as an element of
the abelian group $\Z_2\times\Z_2\times\Z_2$, see \cite{MO}.
The usual product of quaternions then
satisfies the condition
\begin{equation}
\label{ComPr}
a\,{}b=(-1)^{\left\langle\bar{a},\bar{b}\right\rangle}\,b\,{}a,
\end{equation}
where $a,b$ are homogeneous (i.e., proportional to the basis elements) and where
$\langle\,,\,\rangle$ is the usual \textit{scalar product} of 3-vectors
\footnote{
Indeed, $\left\langle\bar{i},\,\bar{j}\right\rangle=1$ and similarly for $k$,
so that $i,j$ and $k$ anticommute with each other.
But,
$\left\langle\bar{i},\,\bar{i}\right\rangle=0$,
so that $i,j,k$ commute with themselves.
}.

\subsection*{The purpose of this talk}

Motivated by the above example, we define the notion of
$\G$-commutative algebra for an abelian group $\G$.
We compare our definition with other known versions of
generalized commutativity: the classical supercommutativity,
as wekk as with more recent notions of
$\b$-commutativity and of twisted commutative algebra.

We formulate a recent classification result
characterizing the Clifford algebras
as the only simple associative $\G$-commutative algebras (see \cite{MO1}).

Our main goal is to attract an interest to this subject
of a wide number of mathematicians working in different areas.
We formulate an open problem and some perspectives of
further development.
The subject is on the crossroads of algebra and differential geometry.
It is closely related to the theory of Lie superalgebras
and geometry of supermanifolds.

\section{What is a commutative algebra?}

This question is not as naive as it seems to be.

\subsection*{The definition.}

The way we understand commutativity is as follows.

\begin{defi}
\label{MDef}
Let $\left(\G,+\right)$ be an abelian group and
\begin{equation}
\label{BiMp}
\langle\,,\,\rangle:\G\times\G\to\Z_2
\end{equation}
a bilinear map.
An algebra $\A$ will be called a $\G$-graded commutative
(or $\G$-commutative) if $\A$ is $\G$-graded:
$$
\A=\bigoplus_{\g\in\G}\A_\g,
\qquad
\A_\g\cdot\A_{\g'}\subset\A_{\g+\g'}
$$
and the condition (\ref{ComPr}) is satisfied
for all homogeneous elements $a,b\in\A$
of degree $\bar{a},\bar{b}$, respectively.
\end{defi}

Let us compare this definition with
other ways to understand commutativity.

\subsection*{Classical supercommutativity.}

The classical notion of supercommutativity is a particular case
of Definition \ref{MDef}
corresponding to $\G=\Z_2$ and $\langle\,,\,\rangle$ the standard product.
A $\Z_2$-commutative algebra is called a
\textit{supercommutative algebra}.

The main examples of associative supercommutative algebras
are the algebras of functions on supermanifolds.
Note that these algebras cannot be simple
(i.e., they always contain a non-trivial proper ideal).

The classical notion of supercommutativity is too rigid for our
purpose.

\subsection*{The notion of $\b$-commutative algebra.}

A more general notion of commutativity
has been considered recently.
Instead of the bilinear map (\ref{BiMp}) with values in $\Z_2$, one
considers an arbitrary bilinear function $\b:\G\times\G\to\bbK$
(a bicharacter) and assume the condition
$$
a\,{}b=\b(\bar{a},\bar{b})\,b\,{}a.
$$
If $\b$ is symmetric, then $\A$ is called $\b$-commutative.

Simple $\b$-commutative algebras were
studied and classified in \cite{BMZ}.
This notion is less restrictive than Definition \ref{MDef}.

Let us mention that
an application of $\b$-commutative algebras to 
deformation quantization is recently proposed in \cite{Phy}.

\subsection*{Twisted commutative algebras.}

A different framework due to S. Majid is related to Hopf algebra viewpoint.
We cite \cite{AM,AM1} as the most relevant references.
Given a commutative associative algebra $\Cm$ and a
function $F:\Cm\times\Cm\to\K^*$,
define a \textit{twisted} algebra structure
$\A=(\Cm, \cdot_F)$, where the new product is given by
\begin{equation}
\label{Twi}
a\cdot{}_Fb=F(a,b)\,ab.
\end{equation}
It is very easy to check that $\A$ is associative
if and only if $F$ is a 2-cocycle:
\begin{equation}
\label{Coc}
F(ab,c)\,F(a,b)=F(a,bc)\,F(b,c).
\end{equation}

The relation to the previous definitions is as follows.
Consider a function $F$ of the form
$F(a,b)=q^{f(a,b)}$, where $q\in\K^*$.
If $f$ is bilinear
(i.e., $f(a+b,c)=f(a,c)+f(b,c)$ and similarly on the second argument)
then $F$ obviously satisfies (\ref{Coc}).
If, in addition, the algebra $\Cm$ is $\G$-graded and $f$ depends only on the grading,
i.e., $f(a,b)=f(\bar{a},\bar{b})$, then $\A$ is a $\b$-commutative
algebra structure with
$\b(\bar{a},\bar{b})=q^{f(\bar{a},\bar{b})-f(\bar{b},\bar{a})}$.

\section{A classification result}

All the algebras and vector spaces we will consider
are defined over the ground field $\bbK=\C$ or $\R$.

\subsection*{Universality of $\left(\Z_2\right)^n$-grading.}

It was proved in \cite{MO1} that $\G=\left(\Z_2\right)^n$
is the only group relevant for the notion of $\G$-commutative algebra.
Furthermore, one can always assume that the bilinear map~$\langle\,,\,\rangle$
is the usual scalar product.

\begin{thm}
\label{RedThm}
(i)
If the abelian group $\G$ is finitely generated, then
for an arbitrary $\G$-commutative algebra $\A$, there exists $n$ such that $\A$ is
$\left(\Z_2\right)^n$-commutative.

(ii)
The bilinear map~$\langle\,,\,\rangle$
can be chosen as the usual scalar product.
\end{thm}

\noindent
It is important to stress that there is no
additional assumption for the algebra $\A$.
We think that this preliminary result is useful for all kind of
classification problems.

\subsection*{Statement of the problem.}

We are interested in simple $\G$-commutative algebras.
Recall that there are two different ways to understand simplicity.

\begin{enumerate}
\item
An algebra is called \textit{simple} if it has
no proper (two-sided) ideal.

\item
An algebra is called \textit{graded-simple} if it has
no proper (two-sided) ideal which itself is a graded subalgebra.
\end{enumerate}

\noindent
We will essentially use the (most classical) definition (1),
but we will also take into account the
definition (2) which is less restrictive.

In order to obtain a classification result, we assume that:
\begin{itemize}
 \item 
$\dim{\A}<\infty$;
 \item 
$\A$ is associative.
\end{itemize}

The best known examples of a simple finite-dimensional
associative algebras are of course the algebras $\M_n$ of $n\times{}n$-matrices.
One is therefore led to the following natural question.
{\it Is there a $\G$-grading on the algebra $\M_n$
such that this algebra can be viewed as a
$\G$-commutative algebra (for a suitable abelian group $\G$)?}
Let us mention that gradings on the algebras of matrices $\M_n$ and more generally
on associative algebras is an important subject,
see \cite{Bah,Bah1,Eld1} and references therein.
However, the $\G$-commutativity condition has not been studied
thoroughly.

\subsection*{The Frobenius theorem.}

A simple associative and commutative (in the usual sense) algebra
is necessarily a division algebra
\footnote{
The proof is elementary and nice, the reader is encouraged to find it.
}.
The associative division algebras are classified by
(a particular case of) the classical Frobenius theorem.
The result is well-known.

\begin{itemize}
 \item
In the complex case, there is a unique simple commutative associative algebra,
namely $\C$ itself.

 \item
In the real case, there is are exactly two simple commutative associative algebras:
$\R$ and $\C$.
\end{itemize}

The following result can be understood as a
``graded version'' of the Frobenius theorem.

\subsection*{Classification in the associative case.}

Here is the main result of \cite{MO1}.

\begin{thm}
\label{CliffC}
Every  finite-dimensional simple associative $\G$-commutative
algebra over $\C$ or over $\R$ is isomorphic to a Clifford algebra. 
\end{thm}

\noindent
The well-known classification of simple Clifford algebras (cf. \cite{Por})
readily gives a complete list:
\begin{enumerate}
\item
The algebras $\Cl_{2m}(\C)$ 
are the only simple associative
$\G$-commuta\-tive algebras over $\C$.
Note that $\Cl_{2m}(\C)$ is isomorphic to the algebra $\M_{2^m}(\C)$
of complex $2^{m}\times2^{m}$-matrices.

\item
The real Clifford algebras $\Cl_{p,q}$ with $p-q\not=4k+1$
and the algebras $\Cl_{2m}(\C)$ viewed as algebras over $\R$
are the only real simple associative
$\G$-commuta\-tive algebras.
Note that all these Clifford algebras are isomorphic to the algebras
$\M_{2^m}(\R),\M_{2^m}(\C),\M_{2^m}(\bbH)$
\end{enumerate}

\noindent
In particular, Theorem \ref{CliffC} answers the above question.

\begin{cor}
\label{CliffCor}
The algebra $\M_n$
can be realized as $\G$-commutative algebra, if and only if $n=2^m$.
\end{cor}

Let us show here that Clifford algebras are indeed
$\G$-commutative.
We refer to~\cite{MO1} for a complete proof of Theorem \ref{CliffC}.

\subsection*{Clifford algebras as
$\left(\Z_2\right)^{n+1}$-commutative algebras}

Recall that a real Clifford algebra $\Cl_{p,q}$
is an associative algebra with unit $\e$
and $n=p+q$ generators
$\a_1,\ldots,\a_n$ subject to the relations
$$
\a_i\a_j=-\a_j\a_i,
\qquad
\a_i^2=\left\{
\begin{array}{rl}
\e,&1\leq{}i\leq{}p\\[4pt]
-\e,&p<i\leq{}n.
\end{array}
\right.
$$
The complex $n$-generated Clifford algebra $\Cl_n=\Cl_{p,q}\otimes\C$ 
can be defined by the same formul\ae,
but one can always choose the generators in such a way that $\a_i^2=\e$
for all $i$.

Assign the following degree to every basis element (see \cite{MO1}).
\begin{equation}
\label{DegCl}
\begin{array}{rcl}
\overline{\a_1}&=&(1,0,0,\ldots,0,1),\\[4pt]
\overline{\a_2}&=&(0,1,0,\ldots,0,1),\\[4pt]
\cdots&&\\[4pt]
\overline{\a_n}&=&(0,0,0\ldots,1,1).
\end{array}
\end{equation}
One then has $\langle\a_i,\a_j\rangle=1$ so that
the anticommuting generators $\a_i$ and $\a_j$
become commuting in the $\left(\Z_2\right)^{n+1}$-grading sense.
If follows that 
an $n$-generated Clifford algebra is
$\left(\Z_2\right)^{n+1}$-commutative algebra,
the bilinear map~$\langle\,,\,\rangle$ being the usual scalar product.

Albuquerque-Majid \cite{AM1} showed that $\Cl_n$
can actually be viewed as the group algebra
$\K[\left(\Z_2\right)^n]$ twisted by some 2-cocycle.
In particular, $\Cl_n$ is $\b$-commutative over $\left(\Z_2\right)^n$,
but the bilinear map $\b$ is more complicated than the scalar product.

\subsection*{Clifford algebras as twisted group algebras}

Consider the algebra $(\K[\left(\Z_2\right)^n],\cdot_F)$
 twisted by the 2-cocycle $F(a,b)=(-1)^{f(a,b)}$
\begin{equation}
\label{NashCoc}
F(a,b)=(-1)^{\sum_{i>j}a_i\,b_j}.
\end{equation}
The cocycle condition (\ref{Coc}) is obviously satisfied
since $f$ is bilinear.

The cocycle (\ref{NashCoc}) has a nice and simple combinatorial meaning.
The product in $\K[\left(\Z_2\right)^n]$ is a simple addition
of $n$-tuples of $0$ or $1$.
The twisted multiplication has an additional \textit{sign rule}.
Whenever we exchange ``left-to-right''
two units, we put the ``$-$'' sign
\footnote{We encourage the reader to check this.}.
For instance,
$(0,1,0)\cdot(1,0,0)=-(1,1,0)$.

\begin{prop} \cite{AM1}
In the complex case, $\Cl_n\cong \left(\K[\left(\Z_2\right)^n],\cdot_F \right)$.
\end{prop}

To obtain the real Clifford algebra $\Cl_{0,n}$,
we will proceed in a slightly different way.
Consider the \textit{even subgroup}
$\left(\Z_2\right)_0^{n+1}\subset\left(\Z_2\right)^{n+1}$
consisting in $(n+1)$-tuples of $0,1$ with even number of $1$-entries.

\begin{prop}
The real Clifford algebra
$\Cl_{0,n}\cong\left(\K[\left(\Z_2\right)_0^{n+1}],\cdot_F\right)$.
\end{prop}

\begin{ex}
The algebra of quaternions $\bbH$ is the algebra of
even triplets.
For instance, one has:
$$
i\cdot{}j \leftrightarrow
(0,1,1)\cdot(1,0,1)=
(1,1,0) \leftrightarrow k,
$$
since the total number of exchanges is even and
$$
j\cdot{}i \leftrightarrow
(1,0,1) \cdot(0,1,1)=
-(1,1,0)
\leftrightarrow-k,
$$
since the total number of exchanges is odd.
In this way, one recovers the complete multiplication table of $\bbH$.
\end{ex}

\section{From graded algebras to geometry}\label{QSec}

Among a huge number of open problems we selected one
challenging problem that looks particularly promising.

A large class of algebras can be viewed as $\G$-commutative.
This opens a possibility to apply algebraic geometry
in order to define geometric objects.
The key notions to be investigated are that of spectrum of a
$\G$-commutative algebra.
We think that such a theory could in particular bring a new
viewpoint to the Clifford analysis.

It worse noticing that, in the simplest $\Z_2$-commutative case,
the theory is very well developed.
However, even the notion of supermanifold is far of being obvious
and involves a sophisticated technique of algebraic geometry.
This notion is still intensively discussed in the literature.

\bigskip

\noindent \textbf{Acknowledgments}.
The second author would like to thank the organizers of the XXVIII Workshop
Geometrical Methods in Mathematical Physics.

\bigskip



\begin{thebibliography}{99}

\bibitem{AM}
H. Albuquerque, S. Majid, 
{\it Quasialgebra structure of the octonions}, J. Algebra
{\bf 220} (1999), 188--224.

\bibitem{AM1}
H. Albuquerque, S. Majid, 
{\it Clifford algebras obtained by twisting of group algebras},
J. Pure Appl. Algebra  {\bf 171}  (2002), 133--148.

\bibitem{BMZ}
Yu. Bahturin, S. Montgomery, M. Zaicev,
{\it Generalized Lie Solvability of Associative 
Algebras}, in: Proceedings of the International Workshop on 
Groups, Rings, Lie and Hopf Algebras, St. John's, Kluwer, 
Dordrecht, 2003, pp. 1--23.

\bibitem{Bah}
Yu. Bahturin, S. Sehgal, M. Zaicev,
{\it Group gradings on associative algebras}, 
J. Algebra  {\bf 241}  (2001), 677--698.

\bibitem{Bah1}
Yu. Bahturin, S. Sehgal, M. Zaicev,
{\it Finite-dimensional simple graded algebras}, 
Sb. Math., {\bf 199} (2008), 965--983.

\bibitem{Phy}
A. de Goursac, T. Masson, J-C.Wallet
{\it Noncommutative $\varepsilon$-graded connections and application to Moyal space.}
arXiv:0811.3567.

\bibitem{Eld1}
A. Elduque,
{\it  Gradings on symmetric composition algebras},
arXiv:0809.1922.

\bibitem{MO}
S. Morier-Genoud, V. Ovsienko, 
{\it Well, Papa, can you multiply triplets?},
Mathematical Intelligencer, to appear.

\bibitem{MO1}
S. Morier-Genoud, V. Ovsienko, 
{\it Simple graded commutative algebras},
arXiv:0904.2825.

\bibitem{Ovs}
V. Ovsienko,
{\it Lie antialgebras: pr\'emices},
arXiv:0705.1629.

\bibitem{Por}
I.R. Porteous,
Clifford Algebras and the Classical Groups,
Cambridge University Press, 1995.

\end{thebibliography}
\end{document}